\documentclass{aa}  

\usepackage{graphicx}
\usepackage{txfonts}
\usepackage{lipsum}
\usepackage{subcaption}         
\usepackage{lscape}             
\usepackage{placeins}           
\usepackage{xcolor}                                

\begin{document}

   \title{Circular polarimetry of suspect wind-accreting pre-polars II}

   \subtitle{}

%

   \author{Pasi Hakala\inst{1}
        \and Steven Parsons\inst{2}
        \and Gavin Ramsay\inst{3}
        \and Boris T. G\"ansicke\inst{4}
        \and Alex J. Brown\inst{5}
        }

   \institute{Finnish Centre for Astronomy with ESO (FINCA), University of Turku, FI-20014, Finland \\
             \email{pahakala@utu.fi}
            \and Department of Physics and Astronomy, University of Sheffield, Sheffield S3 7RH, UK.
            \and Armagh Observatory and Planetarium, College Hill, Armagh BT61 9DG, UK.
            \and Department of Physics, University of Warwick, Coventry CV4 7AL, UK.
            \and Hamburger Sternwarte, University of Hamburg, Gojenbergsweg 112, 21029 Hamburg, Germany.}

   \date{Received September 30, 20XX}

 
  \abstract
   {The origin of white dwarf magnetic fields is an open question. Furthermore, such fields play a vital role in the evolution of interacting binary stars. Here we present a study of white dwarf fields in so called pre-polars (or low accretion rate polars, LARPs , i.e. magnetic cataclysmic variables, where the mass losing secondary star is not in Roche lobe contact, but the systems experience accretion via stellar wind. Such systems play a crucial role in understanding the magnetic braking and the angular momentum evolution of close binaries.}
   {We aim to identify a set of wind-accreting pre-polars amongst our sample of nine candidate systems. We also attempt to measure the white dwarf magnetic fields and study the accretion geometry associated with wind-accreting pre-polars.} 
   {We have obtained optical circular imaging photopolarimetry and circular spectropolarimetry of the targets. These are used to model the cyclotron emission and to estimate the white dwarf magnetic field strengths.}
   {We find that at least in four out of our nine candidate systems, we can confirm the magnetic nature of the white dwarf. Out of these, One system, ZTF J1737+4013, appears to be an eclipsing polar. Furthermore, ZTF J2220+0721 and ZTF J2353+4153 appear to be very strong candidates for wind-accreting pre-polars (or LARPs), with the fourth, ZTF J0056+4926, showing strong variable $H_{\alpha}$ emission, atypical of pre-polars. However, the emission could be related to the strong activity of the donor star, dominating the optical spectrum. One system, ZTF J0504+2145, is likely a novalike CV. The exact nature of the other systems remains unclear.}
   {} 
   \keywords{accretion --
                cataclysmic variables --
                magnetic fields --
                white dwarfs
               }

   \maketitle
   \nolinenumbers

\section{Introduction}

Cataclysmic variables (CVs) are semi-detached binary systems, where a cool donor, typically a main-sequence star (i.e. the secondary), fills its Roche lobe (see \citet{Warner1995} for an early review). Matter is transferred through the L${1}$ Lagrangian point and accreted by the primary, a white dwarf (WD). In non-magnetic CVs, an accretion disc forms around the primary. However, when the WD possesses a sufficiently strong magnetic field (e.g. $\sim$1MG or stronger), the disc is partially disrupted, and the inner accretion flow is redirected along the magnetic field lines to the surface of the magnetic WD (MWD). For stronger magnetic fields (i.e. $\sim$10MG or higher), the accretion disc is completely disrupted, and the ballistic stream of material from the L${1}$ point is re-directed along the MWD's magnetic field lines, eventually impacting the WD near the magnetic pole(s) (see, e.g., \citealt{Cropper1990}). This configuration leads to a synchronous spin of the MWD with the orbital period of the system, resulting in polars (or AM Her binaries).

The strong magnetic field of the WD, together with the magnetic field of the secondary star also play a significant role in the evolution of these binaries. These systems typically form following a common envelope (CE) phase of binary evolution \citep{Pacz1967}, during which one binary component is engulfed by the outer envelope of the other. This causes a rapid loss of angular momentum, causing the orbit to shrink. As the system evolves, the red giant phase ends, and the binary consists of a main-sequence star and a compact object. These systems typically have orbital periods ranging from hours to several days. Further angular momentum loss, via magnetic braking and gravitational radiation, leads to continued orbital shrinkage \citep{Pacz1967, VZ1981}. Eventually, the low-mass main-sequence star fills its Roche lobe, and as the orbit continues to shrink, mass begins to overflow the L$_{1}$ point and accretes onto the primary, giving rise to a CV if the compact object is a WD, or to an X-ray binary if the compact object is a neutron star or black hole.

\begin{table*}[ht!]
  \begin{tabular}{llllllllr}
    \hline
Source & R.A.(2000) & Dec.(2000) & r mag & Per. (d) & Date & Filter/grism & Mode & Parallax (mas) \\
    \hline\hline
    
ZTF J0056+4926 & 00:56:22.36 & +49:26:25.1& 17.1 & 0.135 & 09-10-2021 &	Grism 10 &	circ.sp.pol. & 5.481 $\pm$ 0.051 \\    

ZTF J0220+6303 & 02:20:04.50 & +63:03:59.0 & 17.9 & 0.099 & 05-10-2021 &	Grism 10 & 	circ.sp.pol. & 11.093 $\pm$ 0.074\\

ZTF J0406+0958 & 04:06:27.20 & +09:58:27.1 & 18.6 & 0.115 & 09-03-2022 &	sky cont. &	circ.im.pol. &  4.755 $\pm$ 0.139\\ 	

 &  &  &  &  & 10-03-2022 &	sky cont. &	circ.im.pol. & \\ 	

ZTF J0504+2145 & 05:04:59.80 & +21:45:54.0 & 18.4 & 0.128 & 07-10-2021 &	Grism 10 & circ.sp.pol. & 0.525 $\pm$ 0.187 \\ 

  &  &  &  & & 16-02-2026 & KG5(TNT) & photometry & \\

ZTF J0753+3203 & 07:53:50.78 & +32:02:21.5 & 18.7 & 0.196 & 09-03-2022 & sky cont. & circ.im.pol. & 1.542 $\pm$ 0.208\\ 

ZTF J1520+3453 & 15:20:52.13 & +34:53:44.2 & 16.6 & 0.288 & 10-03-2022 & Grism 10 & circ.sp.pol.& 6.801 $\pm$ 0.042 \\

ZTF J1737+4013 & 17:37:48.34 & +40:43:04.4 & 18.0 & 0.094 & 09-03-2022 & Grism 10 & circ.sp.pol. & 0.982 $\pm$ 0.321 \\

  &  &  &  & & 09-03-2022 & sky cont. & circ.im.pol. & \\

  &  &  &  & & 21-05-2022 & sky cont. & circ.im.pol.  & \\	

ZTF J2220+0721 & 22:20:07.46 & +07:21:29.5 & 18.8 & 0.140 & 09-10-2021 & Grism 10 & circ.sp.pol. & 3.457 $\pm$ 0.230 \\

ZTF J2353+4153 & 23:53:54.96 & +41:53:04.6 & 18.8 & 0.174 & 05-10-2021 & Grism 10 & circ.sp.pol. & 2.832 $\pm$ 0.209 \\

 &  &  &  & & 05-10-2021 & sky cont. & circ.im.pol. & \\

\hline
  \end{tabular}
  \caption{The Observing log with basic information of the targets. The periods come from \protect\url{https://doi.org/10.5281/zenodo.15007293} \citep{zenodopaper} and \citet{2020ApJS..249...18C} for ZTF J0056+4926. Parallaxes are from the Gaia DR3 archive \citep{2016A&A...595A...1G,2023A&A...674A...1G}.}
  \label{table1}
\end{table*}

In Low Accretion Rate Polars (LARPs; \citealt{Schwope2002}), the secondary does not fully fill its Roche lobe, yet magnetic accretion onto the WD is still observed, as evidenced by cyclotron emission from the hot spots on the WD surface. These spots are indicative of a strong magnetic field that channels accretion onto the magnetic poles of the WD. Since there is no Roche lobe overflow, it is suggested that accretion occurs via the stellar wind of the secondary, which is captured by the WD’s magnetosphere (i.e. the magnetic siphon process \citet{Webbink2005_LARP_model}). Such systems have also been referred to as pre-polars (PREPs; \citealt{Schwope2009}), following the recognition that most of these systems do not contain a Roche lobe-filling donor star and that they might have never been in contact with their Roche lobe yet. 

However, according to \cite{Schreiber2021_magWD_binaries} another pathway to a LARP is possible. In this alternative scenario, the detached post common envelope WD+MS binary first becomes a "normal" (i.e. non-magnetic) CV as the system loses angular momentum chiefly by means of magnetic breaking \citet{VZ1981} and the donor star fills its Roche lobe. This is followed by spin-up of the WD rotation due to the accreted matter. If the WD has a crystallised core at this point, the fast rotation generates, via a dynamo process, a strong magnetic field, which eventually connects with that of the donor star. This connection then transfers the gained angular momentum of the WD into the binary orbit, leading to the donor detaching from its Roche lobe and the WD is spin synchronised to the orbital period. There are also other alternatives for the WD magnetic field emergence at later stage. \citet{2022ApJ...935L..12B} have suggested that the initial WD magnetic field generated during the early stages of the WD formation could emerge as a surface magnetic field at the later stage of WD cooling.  After this, the accretion continues as magnetic accretion via stellar wind of the donor onto the magnetic poles of the WD, which the observer detects as a LARP (or pre-polar). Recently \citet{2026MNRAS.547ag521P} have reported a system, where the WD magnetic field might have emerged only after the binary evolutionary track has passed the period minimum. This opens a potentially viable explanation for the observed lack of Roche lobe filling, post period minimum magnetic systems as suggested by \citet{2023A&A...679L...8S}.  

The current known population of LARPs contains 33 systems vs. the total number of "normal" polars of 242 \citep{Schwope2025_PolarCat}. LARPs dominate the aggregate population at orbital periods above the period gap (i.e. >3h) and
they typically show higher WD magnetic field strengths, which is likely at least partially due to a bias resulting from the main discovery method of LARPs (i.e. detecting strong cyclotron humps in the optical band) that favour strong magnetic field systems \citep{Schmidt2005_prePolars}.

Since "normal" polars can exist in both high and low states, and spend $\sim$ 50\% of time in low state \citep{Ramsay2004_PolarsLowState}, it can be difficult to distinguish the LARPs from low accretion state polars. However, there are several recognised observational characteristics that separate the LARPs from traditional polars. These include the lack of accretion stream emission, both in continuum and emission lines and the significantly lower temperatures of the X-ray emitting region \citep{Szkody2004_SDSS_LARP,Vogel2007_WXLMi,Vogel2011_HS0922}. Strong and narrow cyclotron humps in the optical spectrum \citep{Reimers1999} that are also clearly circularly polarised \citep{Schmidt2005_prePolars,Hakala2022} leading to peculiar optical colours, that have been used to pinpoint new candidate systems \citep{vanRoestel2025}. Finally, there is apparent lack of strong flickering \citep{Hakala2022} observed in the optical light curves of polars and almost all other accreting systems at different scales.

This paper is a continuation of \citealt{Hakala2022} (Paper I from now on), where we presented the results from our first polarimetric campaign of pre-polar and/or wind-accreting polar candidates and also adds to the sample identified recently by \citet{vanRoestel2025}. The selection of candidates for this project comes from ZTF-survey, concentrating mostly on eclipsing binary stars containing a WD \citep{zenodopaper}, that appear to show extraordinary light curve modulation profiles and/or colours. Also some of them have shown very fast eclipse ingresses in the subsequent, yet unpublished high speed photometry, suggestive of very compact accretion region(s) on the WD surface. One target, ZTF J0056+4926, comes from \citet{2020ApJS..249...18C}. 

We will proceed by first briefly describing the observations, the setup and reductions, which are more thoroughly explained in Paper I. We will then continue by describing the data and analysis of the results source by source. Finally, we will sum up our findings and discuss our results. 

\begin{figure}
  \begin{center}
\hspace{-5mm}  
\vspace{0mm}
  \includegraphics[width=0.51\textwidth]{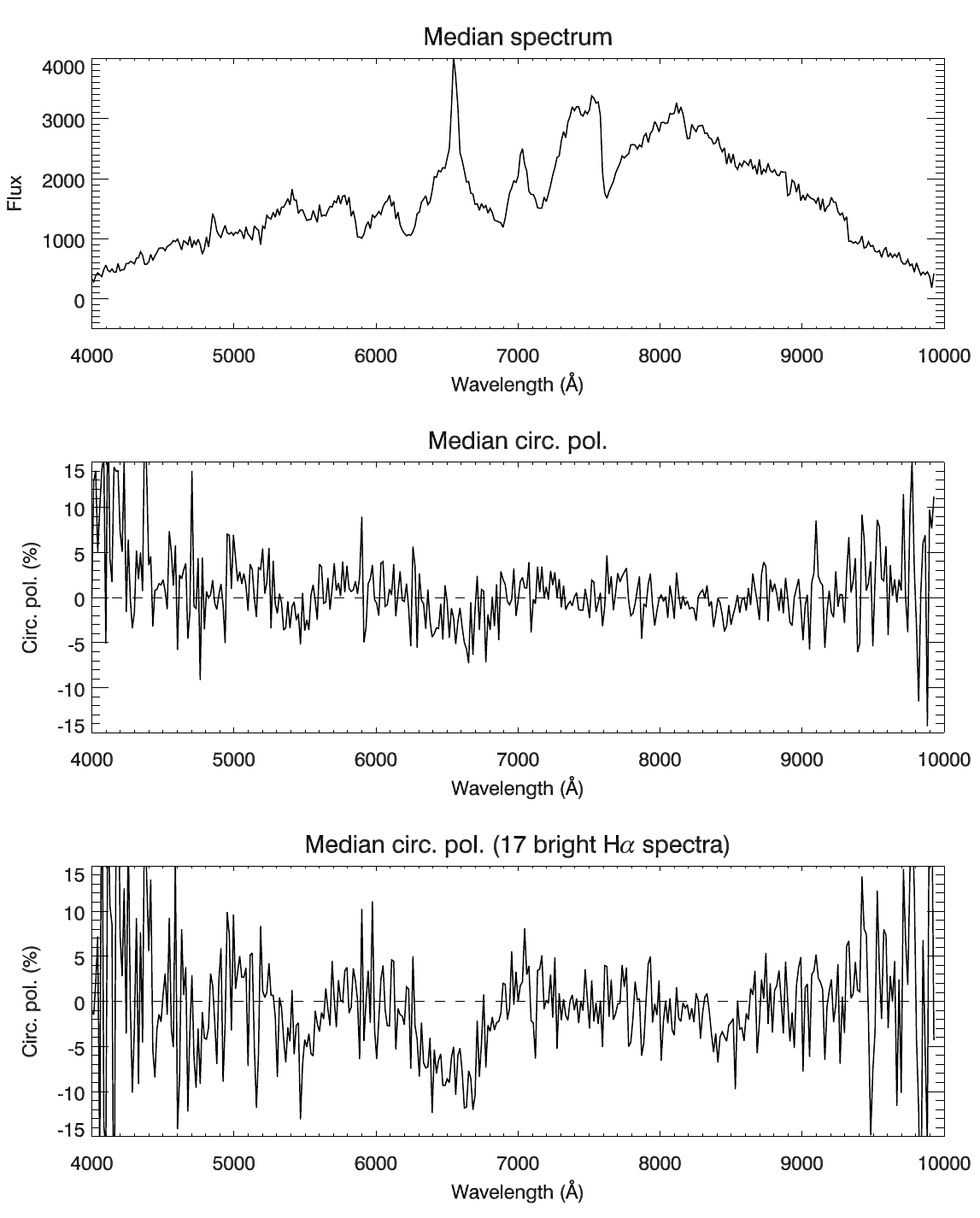}    
\vspace{0mm}
  \caption{Circular spectropolarimetry of ZTF J0056+4926.}
    \label{fig0056spol}
    \end{center}
\end{figure}

\begin{figure}
  \begin{center}
  \vspace{0mm}
  \includegraphics[width=0.5\textwidth,angle=0]{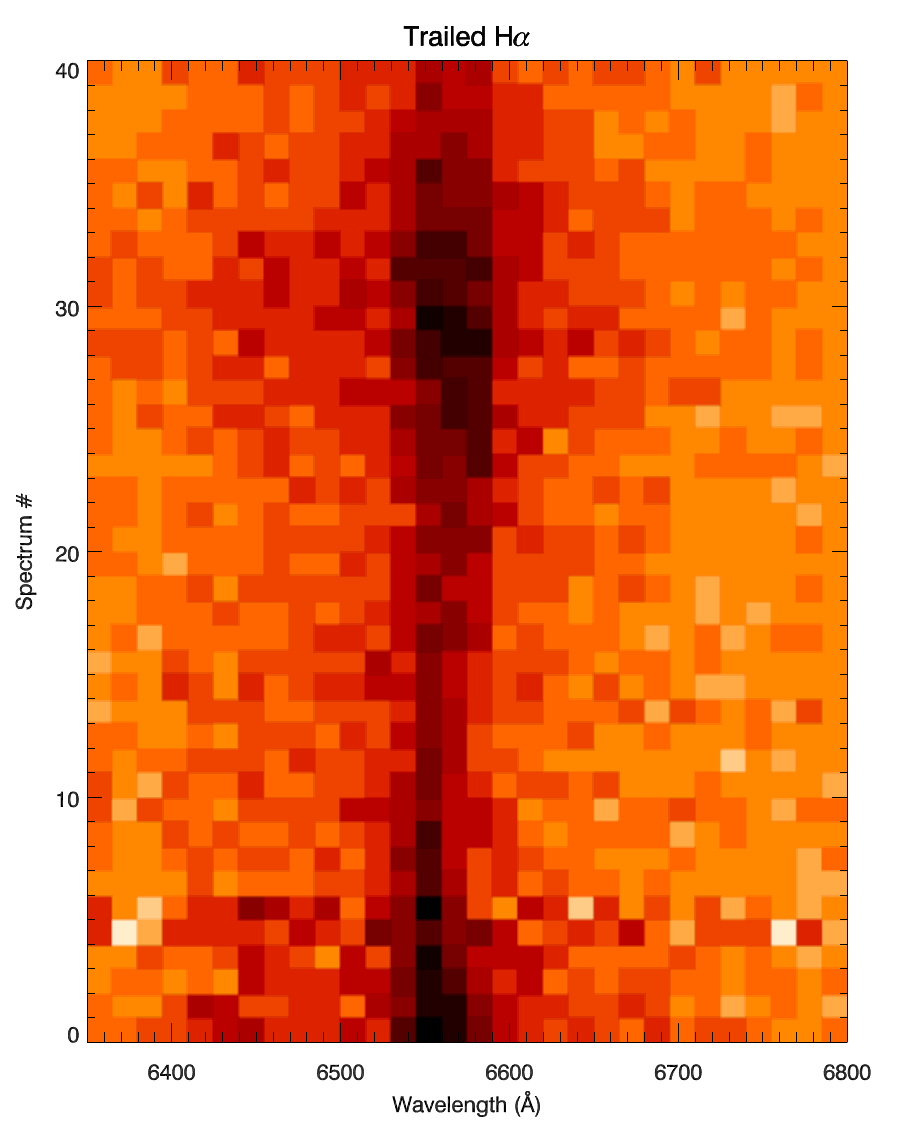}
\vspace{-3mm}
  \caption{The trailed spectrogram of ZTF J0056+4926 in $H_{\alpha}$ covering just over one orbital cycle. Clear
  changes in the line profile and strength are observed.}
    \label{fig0056trail}
    \end{center}
\end{figure}

\begin{figure}
\begin{center}
\hspace{-5mm}
\includegraphics[width=0.5\textwidth]{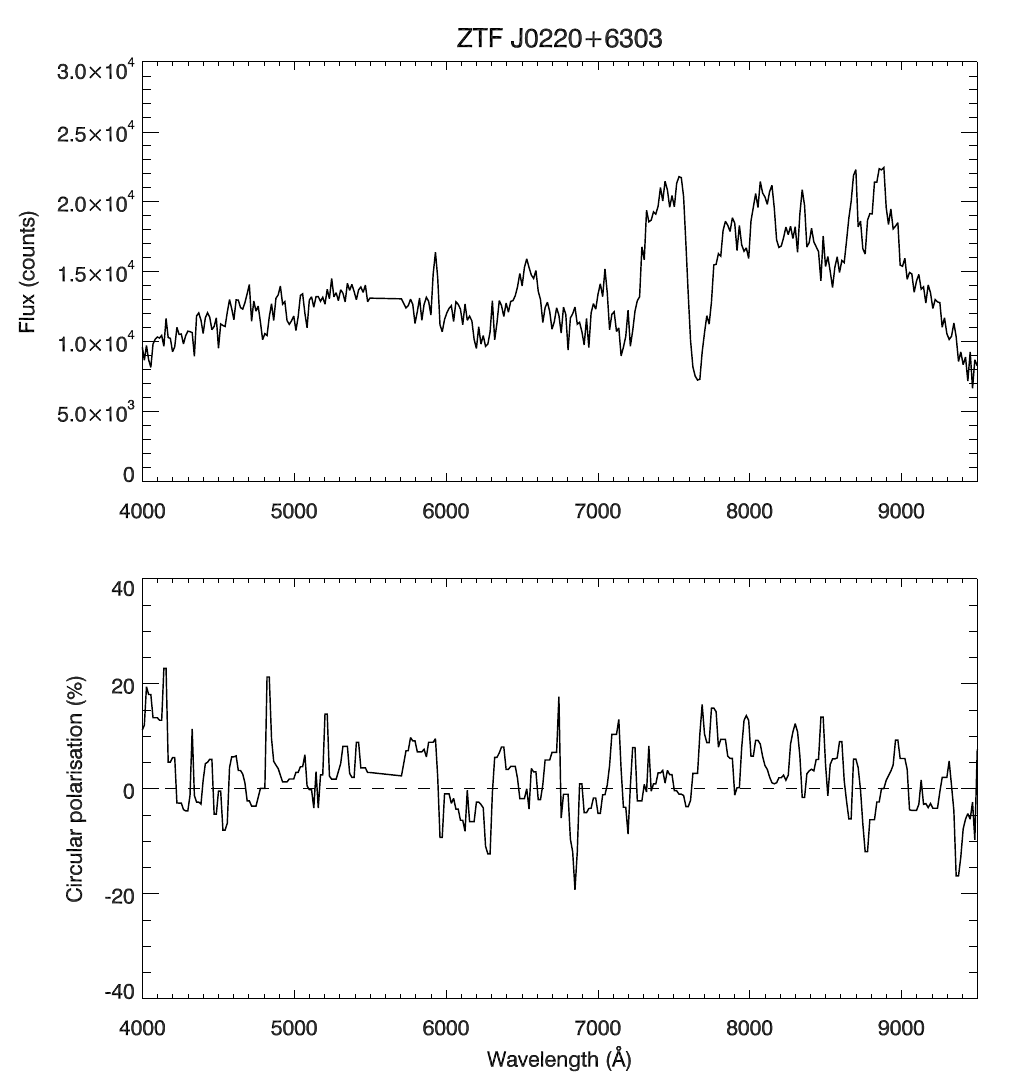}
\vspace{0mm}
\caption{Circular spectropolarimetry of ZTF J0220+6303. the spectrum is dominated by a late type star in the red end and a blue source, possibly showing weak
Balmer absorption lines, in the blue end. The red end of the spectrum (>7000\AA) shows tentative evidence for negative circular polarisation.}
\label{fig0220}
\end{center}
\end{figure}

\begin{figure}
\begin{center}
\hspace{-5mm}
\includegraphics[width=0.51\textwidth]{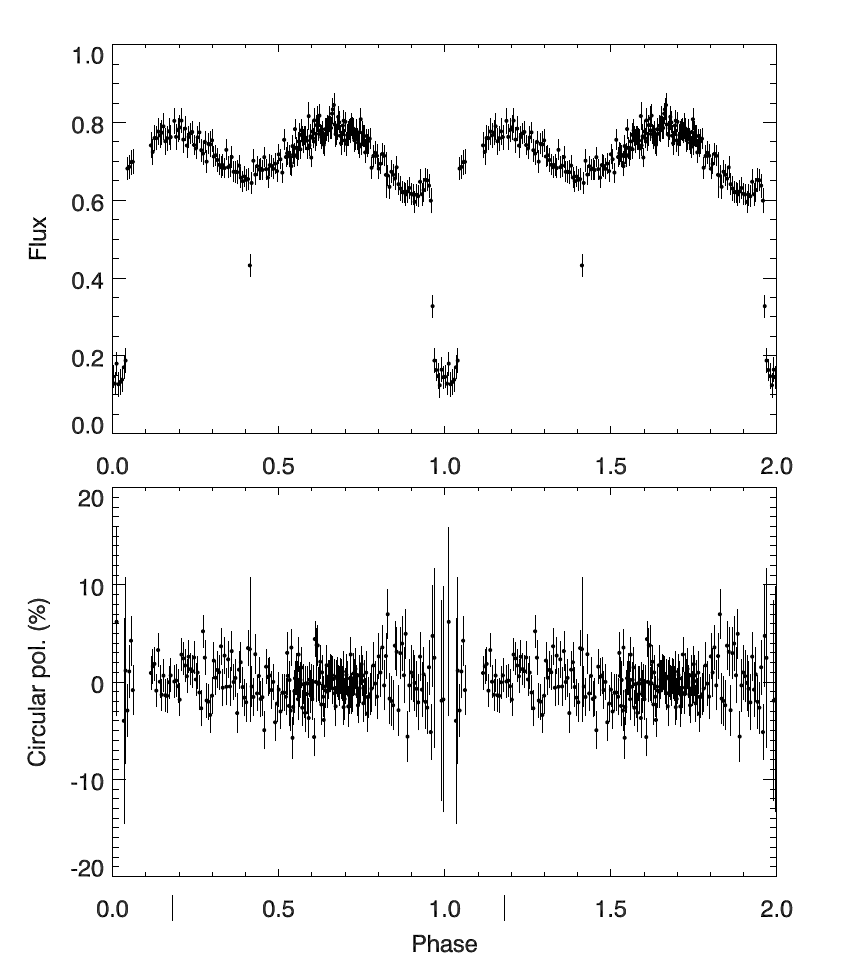}
\vspace{0mm}
\caption{Circular imaging photopolarimetry of ZTF J0406+0958. Data from two nights is folded over the period without binning. There is marginal evidence for modulated circular polarisation over the orbital period.}
\label{fig0406}
\end{center}
\end{figure}

\begin{figure}
\begin{center}
\hspace{-5mm}
\includegraphics[width=0.51\textwidth]{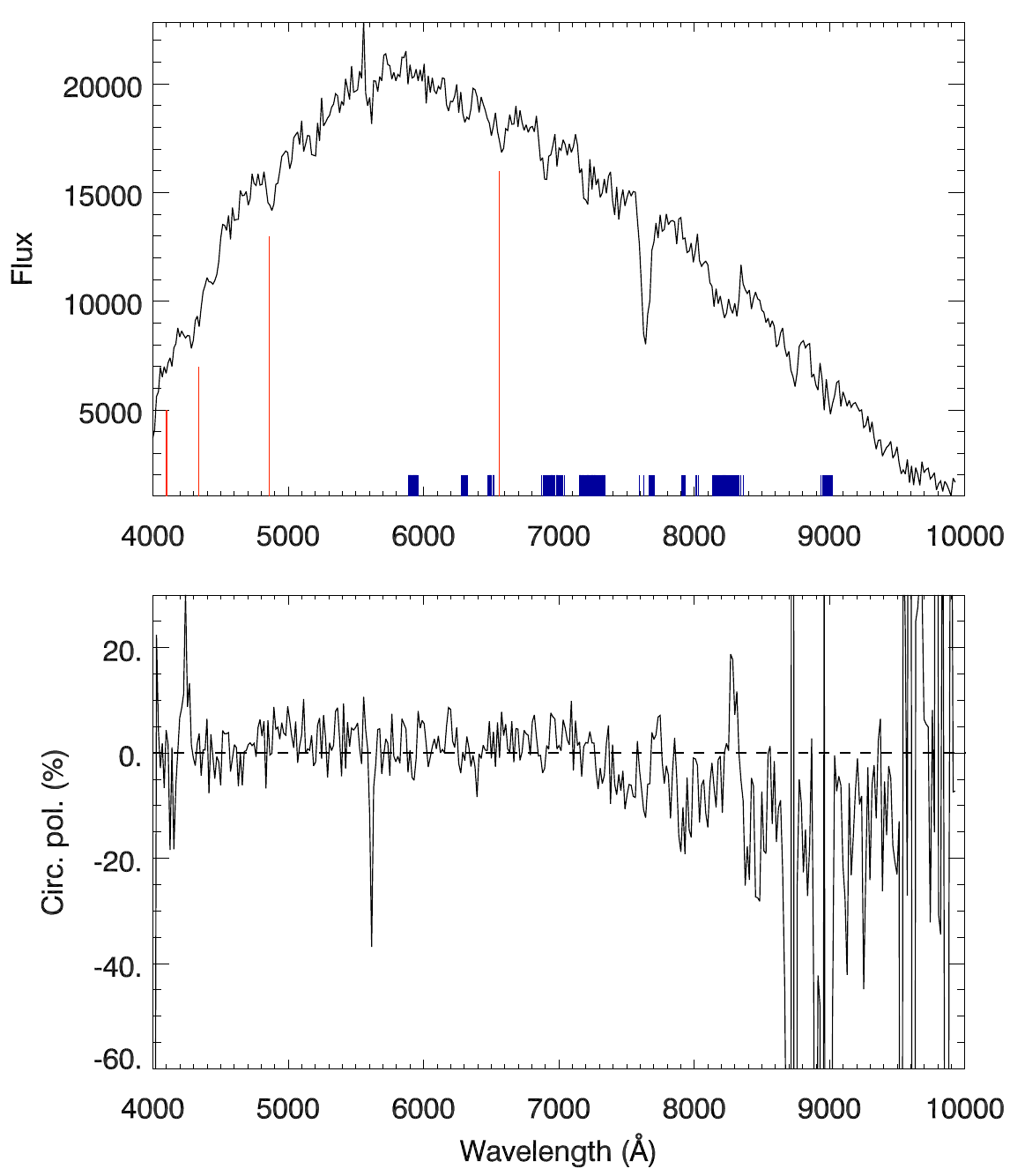}
\vspace{0mm}
\caption{Circular spectropolarimetry of ZTF J0504+2145. the spectrum is dominated by a blue source, showing weak Balmer absorption lines (marked in red). The main telluric features are marked by dark blue blocks in the bottom. The red end of the spectrum (>7000\AA) shows evidence for negative circular polarisation}.
\label{fig0504}
\end{center}
\end{figure}

\begin{figure}
\begin{center}
\hspace{-2mm}
\includegraphics[width=0.49\textwidth]{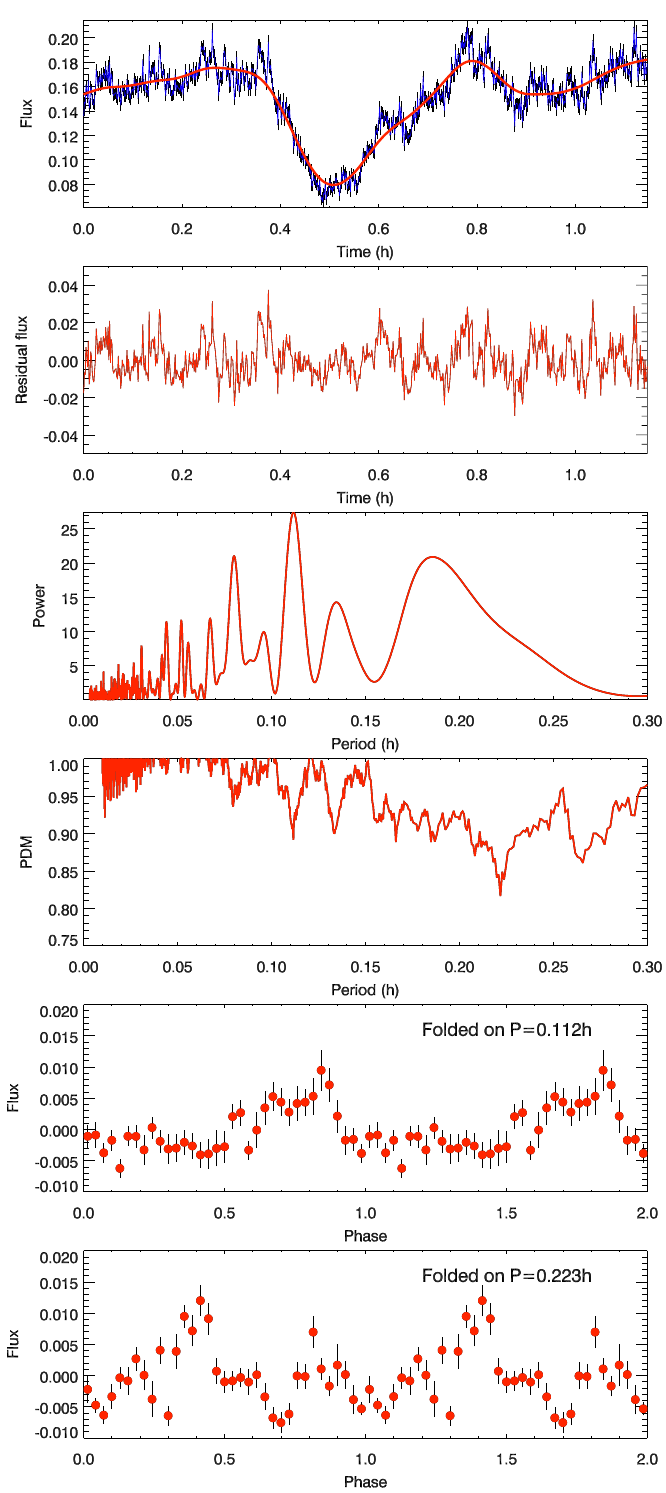}
\vspace{0mm}
\caption{ULTRASPEC photometry of ZTF J0504+2145 in KG5 filter around the eclipse phase, together with the period analysis. See text for details on subpanels.}
\label{fig0504_uspec}
\end{center}
\end{figure}

\begin{figure}
\begin{center}
\hspace{-10mm}  
\includegraphics[width=0.55\textwidth]{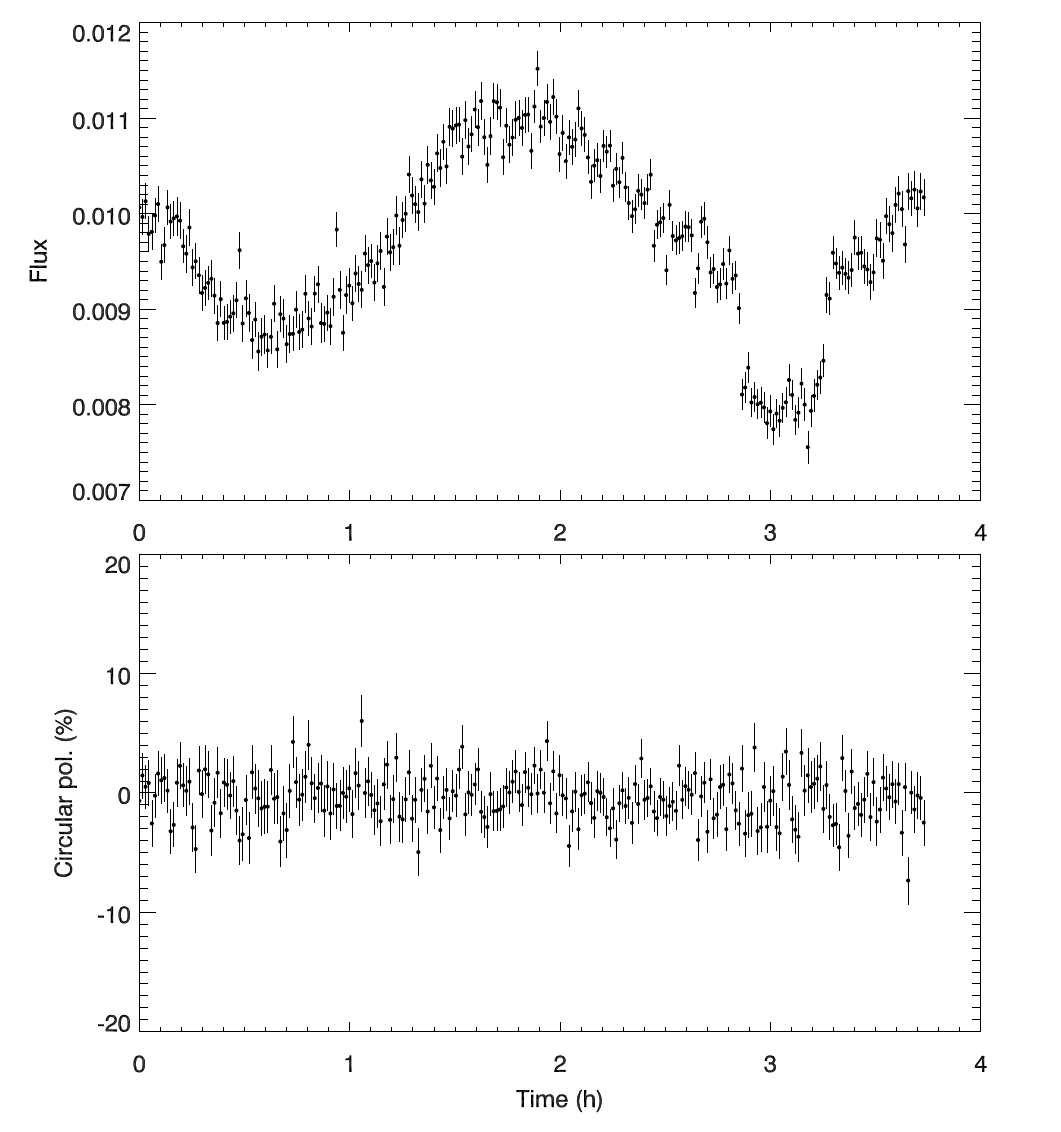}    
\vspace{-3mm}
  \caption{Circular sky contrast band (4000-8000\AA) imaging photopolarimetry of ZTF J0753+3203. }
    \label{fig0753}
    \end{center}
\end{figure}

\begin{figure}
  \begin{center}
  \vspace{0mm}
  \hspace{0.0cm}
  \includegraphics[width=0.5\textwidth]{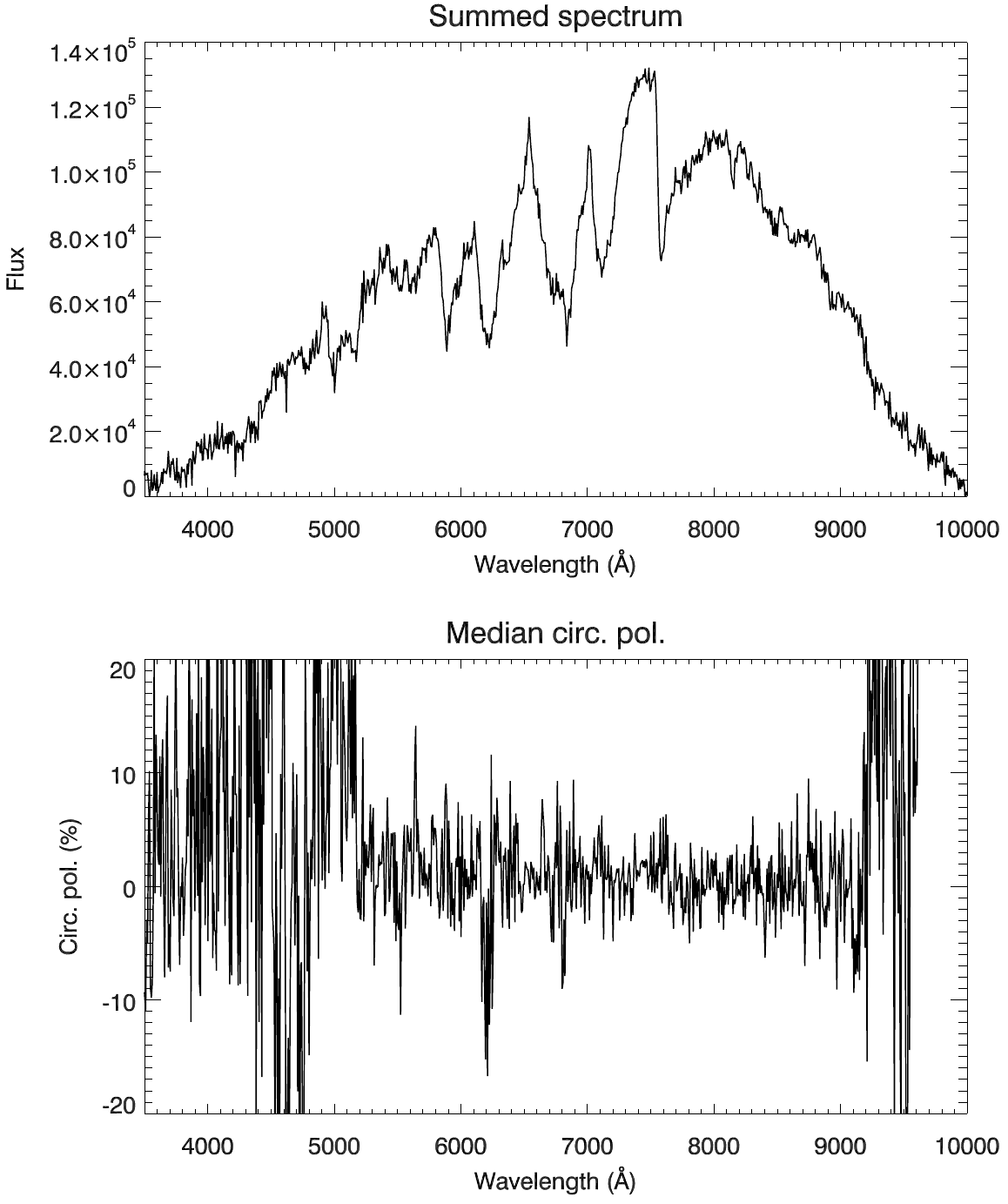}    
\vspace{0mm}
  \caption{Circular spectropolarimetry of ZTF J1520+3453.}
    \label{fig1520}
    \end{center}
\end{figure}

\begin{figure*}
  \begin{center}
  \vspace{0mm}
  \hspace{-5mm}
  \includegraphics[width=0.5\textwidth]{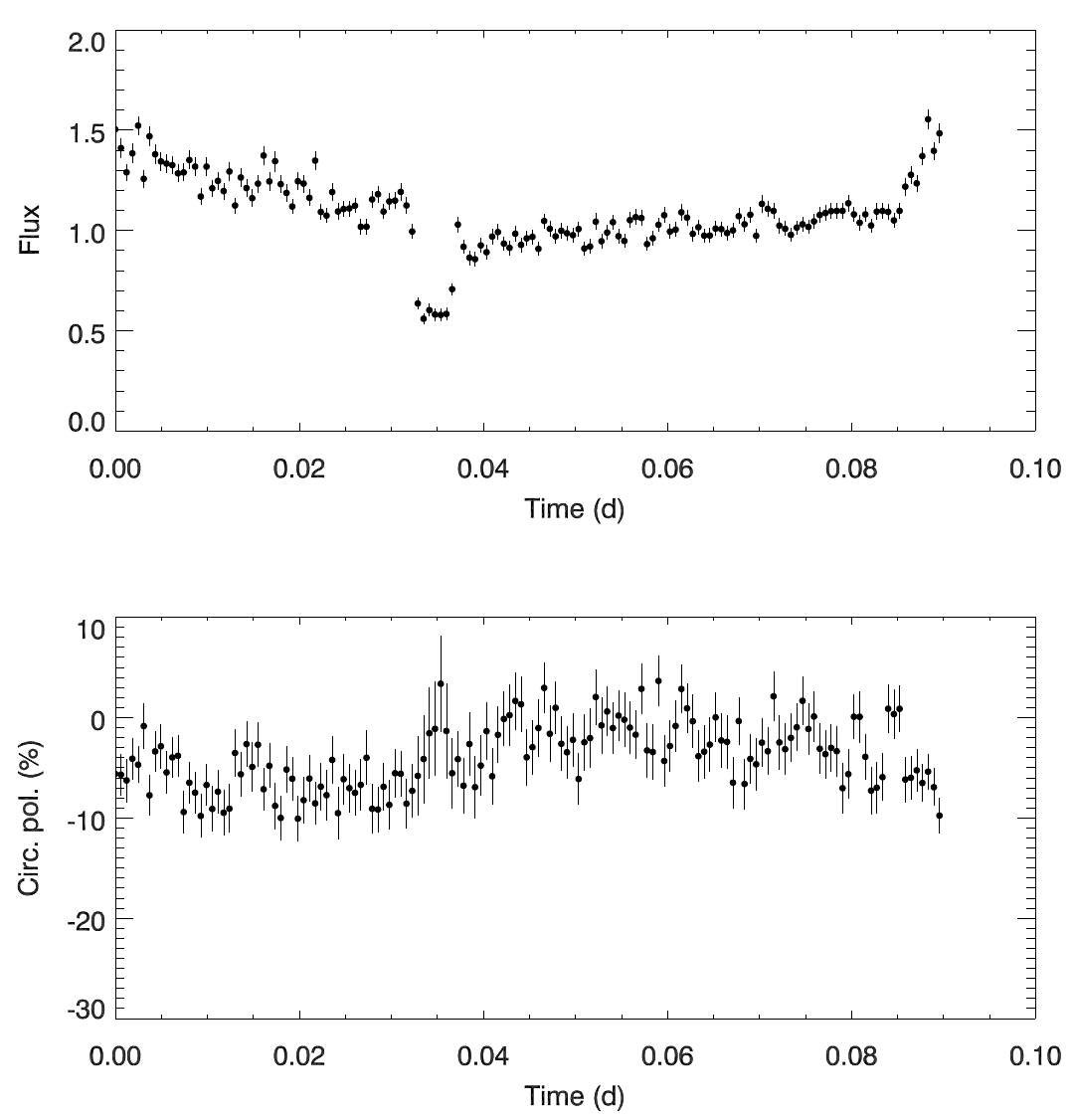} \hspace{-5mm} 
  \includegraphics[width=0.5\textwidth,angle=0]{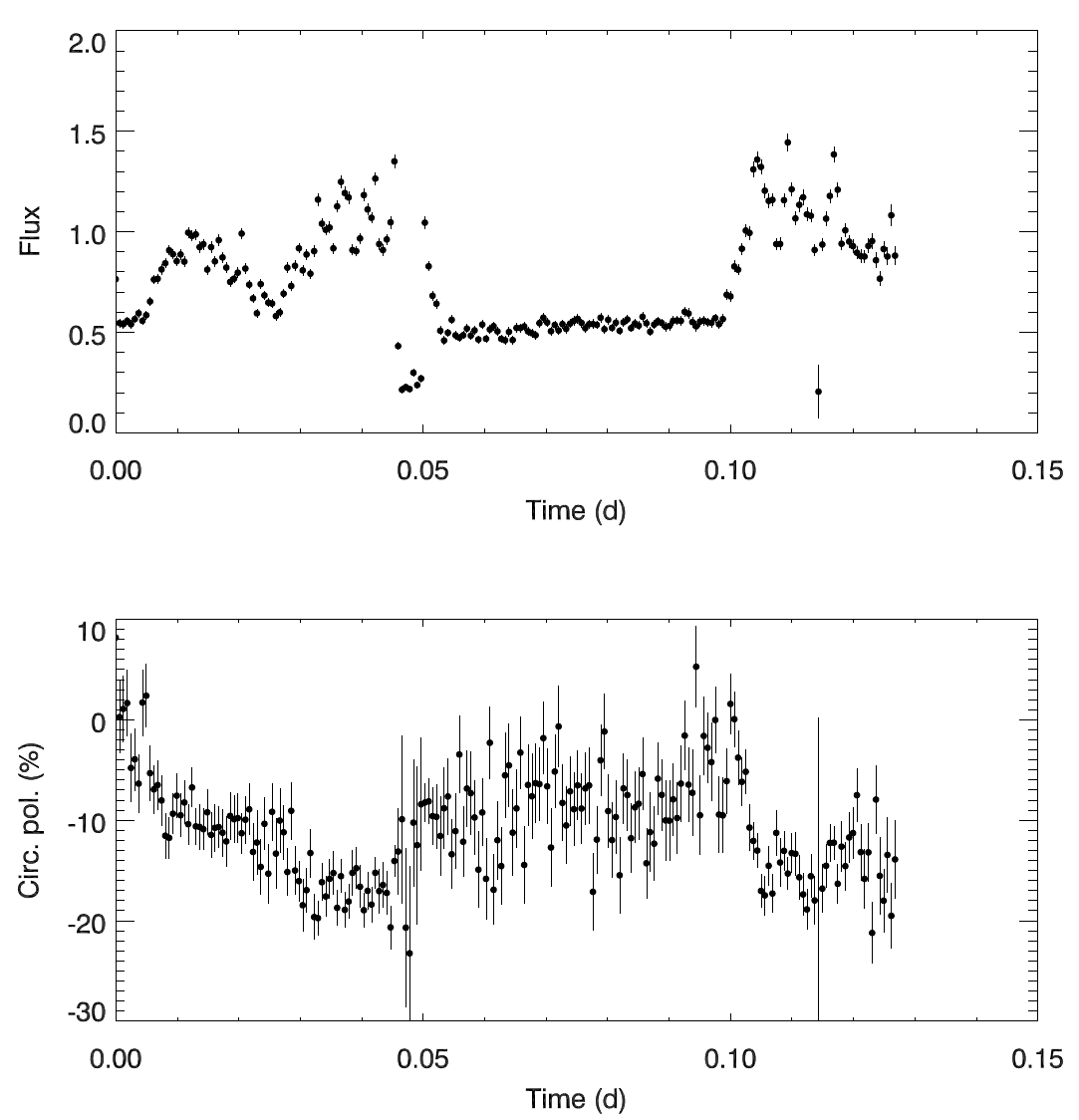}    
\vspace{0mm}
  \caption{Circular photopolarimetry of ZTF J1737+4013. The first (left) and second (right) epochs are shown. The first epoch covers the 0.094d orbital period. The second is longer. The system is in more active state during the second epoch, as evidenced by the increase in circular polarisation and the photometric modulation over the orbital period.}
    \label{fig1737}
    \end{center}
\end{figure*}


\begin{figure}
  \begin{center}
  \vspace{0mm}
  \hspace{-12mm}
  \includegraphics[width=0.54\textwidth]{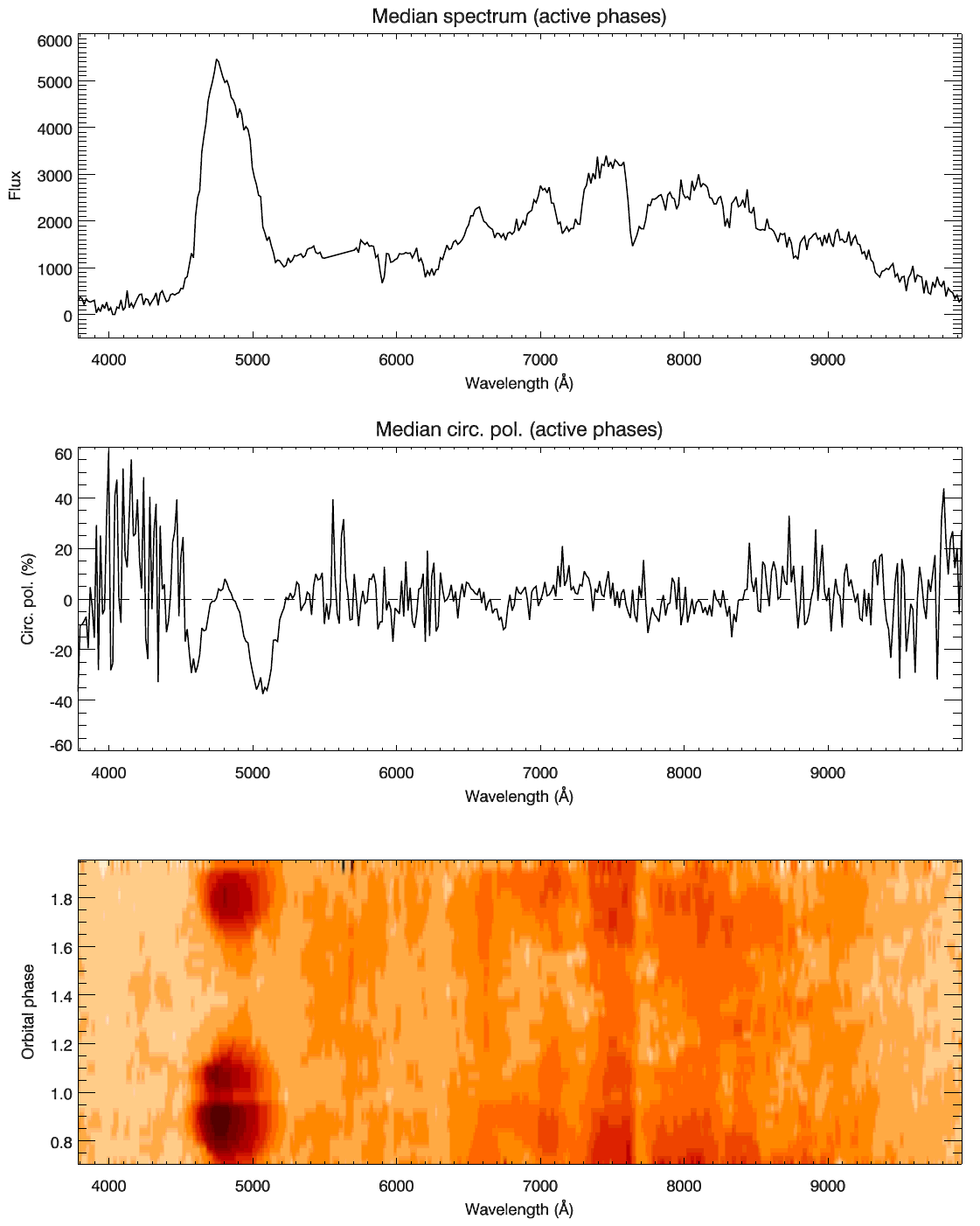}    
\vspace{0mm}
  \caption{Circular spectropolarimetry of ZTF J2220+0721. The top plot shows the median spectrum from the orbital phases, where the accreting pole is visible. The middle plot shows the circularly polarised spectrum from the same phases. The bottom plot shows the flux spectra as a function of orbital phase.}
    \label{fig2220}
    \end{center}
\end{figure}

\begin{figure}
  \begin{center}
  \vspace{0mm}
  \hspace{-6mm}
  \includegraphics[width=0.54\textwidth]{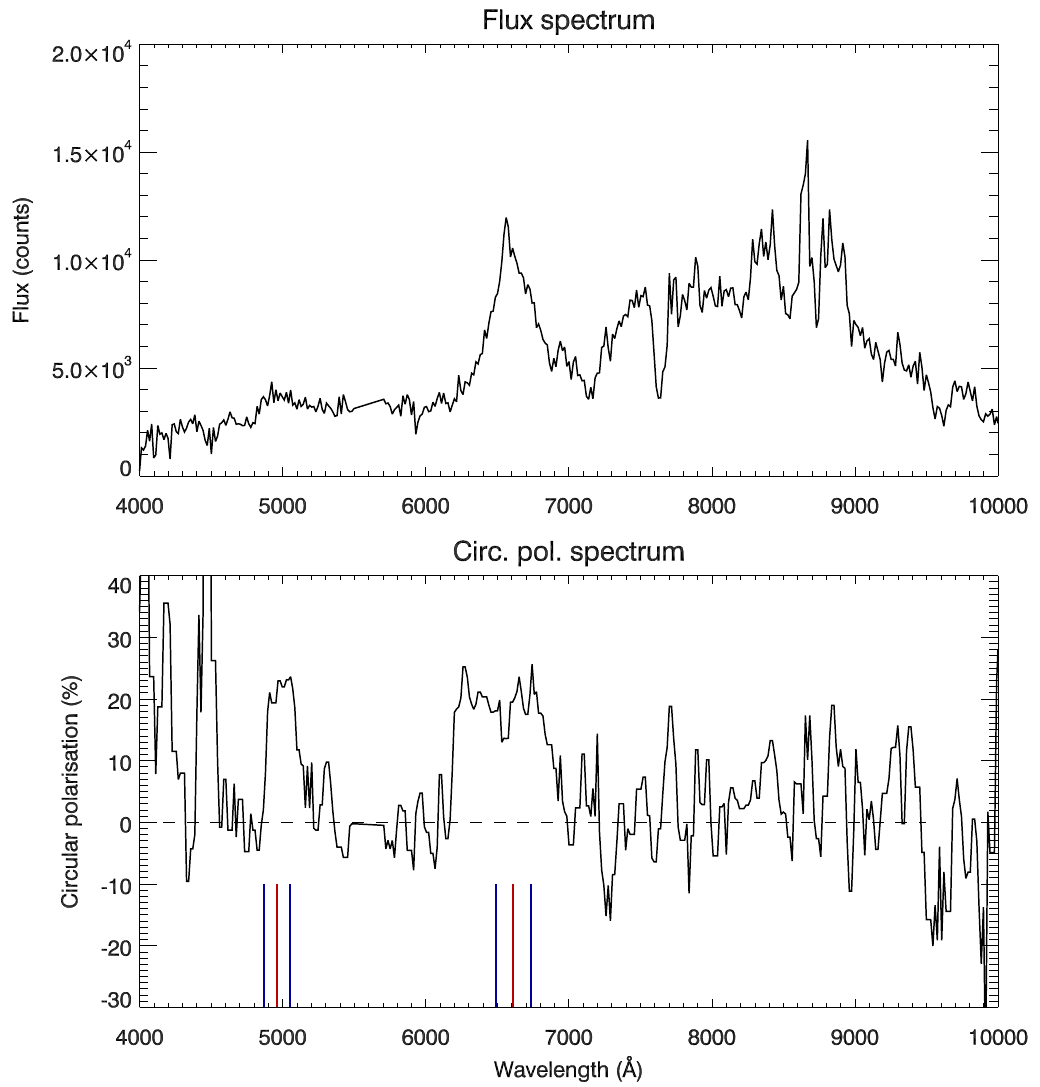}    
\vspace{0mm}
  \caption{Circular spectropolarimetry of ZTF J2353+4153. The top plot shows the flux spectrum. The bottom plot shows the circularly polarised spectrum. We have marked the best fitting cyclotron harmonics (3rd and 4th) from a 54MG cyclotron model with red vertical lines. The blue vertical lines indicate the $\pm$1MG range.}
    \label{fig2353spol}
    \end{center}
\end{figure}


\begin{figure}
  \begin{center}
  \vspace{0mm}
  \hspace{-12mm}
  \includegraphics[width=0.54\textwidth]{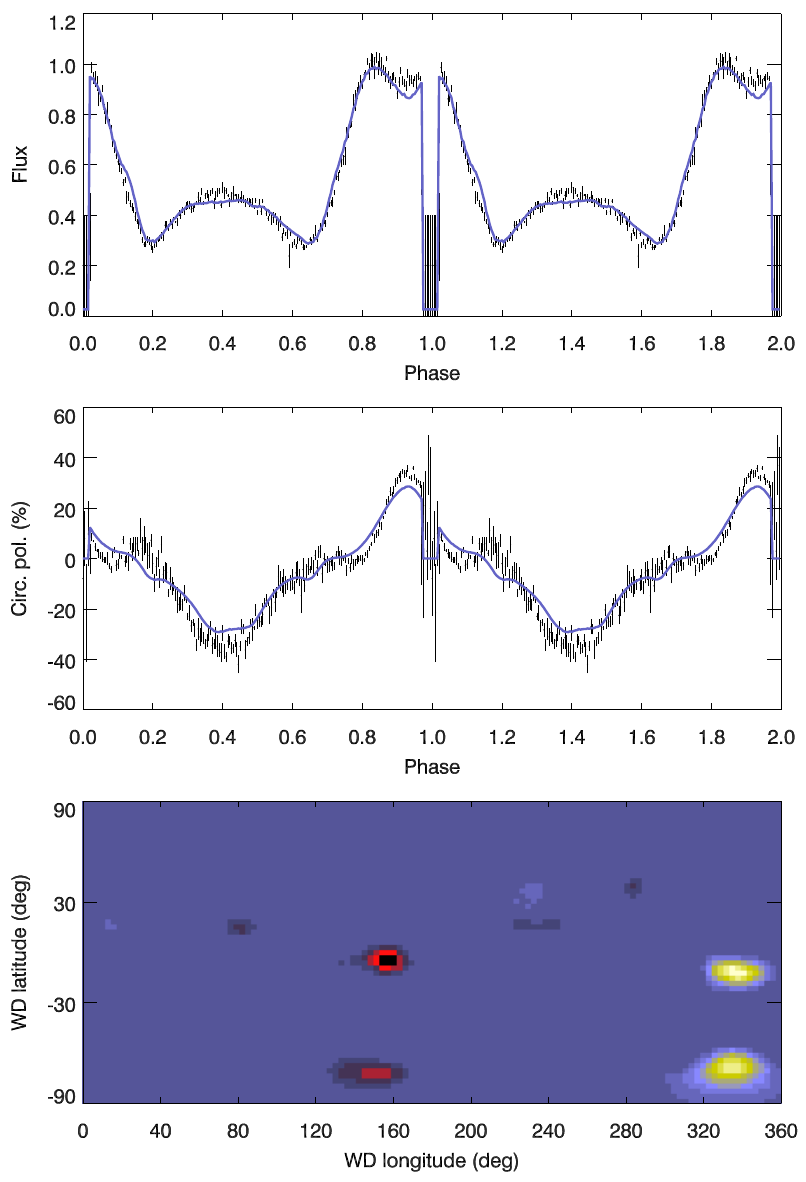}    
\vspace{0mm}
  \caption{Circular fast photopolarimetry of ZTF J2353+4153, together with our model fit. The system is eclipsing with two-pole accretion, as indicated by the reversal in sign of the circular polarisation. The bottom panel shows the resulting cyclotron emission map on the WD surface following an inversion of the circular photopolarimetry data. The yellow/green and red/black areas refer to areas of cyclotron emission with positive and negative polarity respectively.}
    \label{ztf2353model}
    \end{center}
\end{figure}


%

\section{Observations}

The observations for this study were primarily conducted during two observing runs in October 2021 and March 2022 at the Nordic Optical Telescope (NOT), with additional data collected in May 2022. Furthermore, additional photometry for one target was obtained from the Thai 2.4m TNT telescope equipped with ULTRASPEC \citep{2014MNRAS.444.4009D} in February 2026 (see Table \ref{table1} for the observing log). The NOT observations were carried out using the ALFOSC (Andalucia Faint Object Spectrograph and Camera), a versatile imager and spectrograph capable of both polarimetric imaging and spectropolarimetry. Brighter sources were observed in circular spectropolarimetric mode, while fainter systems were observed using imaging circular polarimetry in a single passband (sky contrast filter \#92 that effectively covers the Gaia G-band), chosen to maximize the detection probability of cyclotron humps for typical WD magnetic field strengths in polars. The specific observational modes and configurations are detailed in Table 1.

ALFOSC was equipped with the e2V back-illuminated, deep-depletion CCD with a pixel size of 15 $\mu$m (2048 $\times$ 2064 pixels), providing a full field of view (FOV) of 6.4' $\times$ 6.4'. However, a partial window mode was employed for this study, combined with on-chip binning. All spectral observations were carried out using grism \#10. This grism provides a spectral coverage of 3300--11000 \AA. The formal spectral resolution of the grism is $R = 105$ when used with a 1" slit. However, in this work, we employed a 1.8" polarimetric slitlet and additionally applied a binning factor of two in both the spectral and spatial directions, resulting in an effective resolution of $R \sim 50$ for the majority of sources.

For the polarimetric observations, a calcite block was used in conjunction with a rotatable quarter-wave plate, which was rotated in 90$^\circ$ steps. Data were acquired at four different retarder plate angles, allowing for the automatic cancellation of optical system transmission effects in both spectropolarimetric and polarimetric imaging modes. In certain cases, observations were made using only two angles (0$^\circ$ and 90$^\circ$), which provided a higher temporal resolution, beneficial for monitoring variable sources.

All data were subjected to bias subtraction, and the spectra were flat-fielded using exposures from a halogen lamp. The circular polarisation imaging time series data were obtained using autoguiding, with the quarter-wave plate rotating in 90$^\circ$ steps. No flat-field correction was required for this mode.  During imaging polarimetry observations, field stars within the FOV were used to verify both the zero level of circular polarisation and the photometric calibration.

NOT data reduction and optimal spectral extraction were performed using custom IDL scripts, specifically the OPTSPECEXTR package \footnote{J. Harrington, \url{https://physics.ucf.edu/~jh/ast/software/optspecextr-0.3.1/}}. Aperture photometry for imaging observations was carried out using custom scripts based on the IDL astronomy library. The ULTRASPEC data were reduced using the HiPERCAM pipeline \citep{2021MNRAS.507..350D}, following the standard procedure.

The basic properties of the observed sources are also shown in Table. \ref{table1}


\section{Analysis and discussion of individual sources}

\subsection{ZTF J0056+4926}

ZTF J0056+4926 has an orbital period of 0.135 d and was observed in circular polarimetry spectroscopic mode on 09-10-2021. 40 pairs of spectra using grism 10 were acquired, each with a 180 sec exposure time. The observations covered $\sim$1.3 orbital periods. This source shows strong variable H$_\alpha$ emission on top of a late type star spectrum. Circular polarisation is not clearly detected in individual spectra. However, adding them up and restricting the analysis to H$\alpha$ bright spectra reveals two humps, at 6700\AA\ and 5400\AA\, that roughly correspond to the 4th and 5th cyclotron harmonics of a $\sim$40MG field (Fig \ref{fig0056spol}), using the formula by \citet{1993MNRAS.262..285F} (i.e. eqn.1), which gives the central wavelengths of cyclotron emission humps as a function of magnetic pole field strength (in Gauss) for the magnetic field viewing angle of 90$^{\circ}$ and cyclotron temperatures appropriate for polars. It should be noted though, that these do depend also on the viewing angle and the plasma parameter $\Lambda$.

The orbital phase range of the spectra showing stronger H$\alpha$ emission and more prominent cyclotron humps is 0.25-0.75. This coincides with the phase range exhibiting more scatter in the ZTF r band light curves (Appendix A.)
However, this scatter in the phase folded data appears to originate only from the first season of ZTF data. We note that the g and r band ZTF light curves are quite different in shape. The r band matches well with the most prominent circularly polarised hump at 6700\AA\, whilst the centre of the second hump at around 5400\AA\ falls in between the g and r bands. This might explain the light curve difference between the g and r bands.

\begin{equation}
    \lambda_n = \frac{10710}{n} \left (\frac{10^8}{B} \right ) \AA
\end{equation}

The third feature in the source spectrum that looks like a possible cyclotron hump (at 8500\AA\ ) does not quite fit with the same field strength, as the 3rd harmonic should be located at $\sim$8900\AA\ . The trailed spectrum, centred on H$_{\alpha}$, shows that the emission line profile behaviour is more polar like than a disc (Fig \ref{fig0056trail}), i.e. there is no indication for double peaked emission, typical for accretion discs. Furthermore, the line width varies strongly over the orbital period. It is also clear that the line consists of several components - also typical for polars. The spectral resolution is extremely low limiting any further analysis. Given the dominance of the donor star flux in the optical spectrum, the accretion rate must be extremely low. The ZTF light curves (Appendix A) reveal that there are clear changes in the r band folded light curves over time, whilst the g band light curve remains relatively stable. This is in line with the strongest circularly polarised cyclotron hump being located at around 6600\AA. The source does not show any major accretion state changes in the ZTF data. This, together with the very low cyclotron emission level when compared to the donor star makes the source a likely LARP. We do note, however, the presence of strong H$_\alpha$ emission and the variations in the folded r band ZTF light curves that might be better explained by a Roche lobe filling polar.

\subsection{ZTF J0220+6303}

ZTF J0220+6303 has a 0.099 d orbital period and shows an optical spectrum dominated by a late type star above 6500\AA\ (Fig \ref{fig0220}). However, the continuum rises substantially towards the blue end and there is clearly another continuum component present at the wavelengths below 6500\AA. This source showed a very deep eclipse in \citet{Brown2023}. The only possibly identifiable spectral feature in this region is an emission line at around $\sim$5900\AA\, which could be HeI at 5876\AA. Based on the current single spectrum, it is not possible to identify the origin of the blue continuum source. It is likely to be either an accretion disc or WD photosphere. The shape of the ZTF light curve \citep{Brown2023} also implies that this blue emission component is variable on the timescale of the orbital period. There is no measured circular polarisation in the spectrum.

\subsection{ZTF J0406+0958}

This source, also identified as an eclipsing binary in \citet{Brown2023}, has an orbital period of 0.115 d and the circular photopolarimetry shows a possible weak modulation at the orbital period, with eclipses superimposed on 
a double-humped photometric modulation (Fig \ref{fig0406}). The 99.9\% upper limit for the circular polarisation amplitude (excluding the eclipse phases) is 1.6\%. ZTF J0406+0958 first appears as a detached MS+WD binary, with ellipsoidal modulation by the donor star. However, the mid eclipse phase does not coincide with the main flux minimum, but is rather delayed by about 0.1 in phase. This is also evident from the ZTF light curves of \citep{Brown2023} making this a persistent feature. Furthermore, 
what looks like an ellipsoidal modulation in the light curve, cannot actually originate from the secondary, since the M dwarf contribution to the total emission is far too small to produce such a prominent modulation. This is evident from the depth of the eclipses, which limit the fractional
emission from the M dwarf to be $\sim$0.2. This is roughly the same as the full amplitude of modulation outside the eclipses. This further rules out starspots and other donor star related phenomena as a viable origin for the observed photometric modulation.  

There is some evidence from the light curve, that during the eclipse the "rising trend" in the out-of-eclipse light curves is NOT seen. This suggests that the overall modulation could come from the WD, as we first expected (i.e the first assumption was that it could be of cyclotron origin). However, we are not detecting any prominent changes in the circular polarisation over the orbital period. It is worth considering whether spots on the WD surface might be able to cause the modulation.
The origin of the eclipse phase delay remains indeterminate based on all the data obtained so far. 

\subsection{ZTF J0504+2145}

This 0.128 d period source shows an early type, possibly WD, spectrum (Fig \ref{fig0504}). The resolution is not sufficient to allow for spectral classification. There is no indication of the secondary star in the spectrum. The phase folded ZTF light curves (Appendix A) indicate asymmetric orbital modulation in both g and r bands, leading to the inclusion of the source in this programme. Intriguingly, the continuum appears to become significantly and systematically more circularly polarised beyond 7000\AA. However, if this corresponds to a wide fundamental frequency, then the 2nd harmonic should be visible at around 4500\AA. Furthermore, the fundamental cyclotron frequency should be optically thick and not significantly polarised. 

Very recently, we have obtained additional photometry of the system using the 2.4m TNT telescope, together with the ULTRASPEC instrument. These observations were carried out using a wide KG5 filter and reduced using the standard procedure/pipelines available for the instrument. Unfortunately the KG5 filter does not exclusively cover the >7000\AA\ region, where the cyclotron polarisation appears, but also extends further towards the blue. Consequently, the light curves not only sample the polarised flux, but also the blue unpolarised continuum. 

The results of the ULTRASPEC photometry are shown in Fig \ref{fig0504_uspec}. The top plot shows the light curve covering an asymmetric eclipse, together with additional modulation. The corresponding orbital phase range coverage is 0.84 - 0.21. This eclipse light curve strongly suggests that ZTF J0504+2145 is a disc accreting CV. We have fitted the long term behaviour with a regressive spline (the smooth curve in red). The second plot from the top shows the residual light curve (with the spline model subtracted). The variability in this light curve suggests that there could be periodicity present with 0.1-0.15h period. To check for this, we carried out period analysis, the results of which can be found in  Fig \ref{fig0504_uspec}. The 3rd and 4th subpanels  show the Lomb-Scargle \citep{1982ApJ...263..835S} and phase dispersion minimisation \citep{1978ApJ...224..953S} periodograms (using 35 phase bins) of the residual light curve. As the periodograms suggest a very tentative presence of periodicity either at 0.11h or 0.22h period, we have also plotted the data phase folded and binned into 35 phase bins in the lowest two panels. The pulse associated with the 0.112h period has a feasible shape that could arise from the changing visibility of the bright spot on the WD surface as the WD spins. In order to further verify the possible significance of the 0.112h period, we performed a red noise analysis, by fitting the eclipse-subtracted light curve by a second order autoregressive timeseries model (along the lines of   \citet{2011MNRAS.416..644H}). We find out that there is a 21\% chance that the 0.11h period is not real, but produced by red noise (i.e. flickering). However, this limit is probably very much affected by the shortness of our time series, as flickering can mimic periodic variability over a handful of cycles.     Given that there is evidence for circular polarisation in the red end of the spectrum, and that there is some visible evidence for possible periodic variability in the ULTRASPEC light curve, the tentative 0.11h period should be noted down for more extended future studies, hopefully to be carried out in r, i or z bands
The tentative 0.11h period is not evident in the ZTF r band light curves, shown in Appendix A. Those light curves also show the system mostly in high state, apart from a single deep low state.
Such behaviour is typical for VY Scl (novalike) systems \citep{2024MNRAS.535.3035D}. Finally, we note that this system has a Gaia parallax of 0.5245 $\pm$ 0.1873 mas, which puts the system at 1.9 $\pm$ 0.68 kpc. This translates to an absolute r magnitude of 7.0, which is in range for Novalike CVs \citep{2017A&A...604A.107R}, but also for intermediate polars (IP) at 3h period \citep{2023MNRAS.523.3192M}. 

\subsection{ZTF J0753+3203}

This appears to be an eclipsing MS+WD binary with 0.196 d period also exhibiting ellipsoidal modulation due to the tidal effects on the secondary star. The source was included as a candidate LARP as it had shown evidence for some changes in the light curve earlier, but those were likely due to the starspots on the donor star (unpublished, priv.comm.).There is no indication of accretion related features in the light curve and there is no detection of any circular polarisation either (Fig \ref{fig0753}). Also, there are no major changes in the ZTF light curves of this source (Appendix A).

\subsection{ZTF J1520+3453}

We obtained $\sim$2 hrs of circular spectropolarimetry of ZTF J1520+3453, covering 0.29 in orbital phase of the 0.288 d period. The overall spectrum (Fig \ref{fig1520}) shows only the flux spectrum of a late type MS star. Furthermore, there is no indication for any circular polarisation in the spectrum. There is some indication for possible H$_\alpha$ emission, likely to originate from the stellar activity of the fast rotating secondary star. Given the fairly limited phase coverage, we cannot entirely rule out a magnetic CV, as we might have missed the orbital phase range when that magnetic pole would have been visible. Our spectra cover the orbital phase range 0.43-0.63 (ephemeris by S. Parsons, priv. comm.). All of our individual spectra show a similar M dwarf spectrum with little variation and negligible circular polarization. Finally, the ZTF light curves (Appendix A) show almost flat orbital light curve, with just slight asymmetry as the 0.75 phase flux exceeds the 0.25 phase flux, leading to the inclusion of this source in our list. The the WD eclipses as the only prominent feature in the ZTF light curves.

\subsection{ZTF J1737+4013}

This eclipsing system, with an orbital period of 0.094 d is clearly an eclipsing polar. There is only a single spectrum of the source, which is of very low S/N and as such cannot provide any further insight. Thus we are limited to what we can derive from the photopolarimetry. This consists of data from two different epochs i.e 09-03-2022 and 21-05-2022.
As shown in Fig \ref{fig1737}, the source shows eclipses with very fast ingresses and egresses, typically produced by a WD (or a cyclotron region on the WD surface). Both datasets are too short to include more than one eclipse each. The first dataset is not conclusive, but the second set displays a hallmark single pole accretor AM Her light curve, where the cyclotron region is obscured behind the WD roughly 50\% of the orbital cycle, leading to flat light curve during that phase range. Also, the secondary minimum in the light curve is suggestive of cyclotron beaming of the emission. Finally, we detect increased level of circular polarisation during the phase range, when the cyclotron region is visible to the observer. 

It is also worth noting that the source was clearly in different accretion state at the different epochs. The first light curve (Fig \ref{fig1737}) shows the source in low state, where there is only marginal amount of cyclotron emission, judging both by the light curve and the circular polarisation curve. However, during the second observation, the cyclotron emission is much more pronounced. This suggests that the source experiences sizeable changes in the accretion rate, which further suggests that this might be a "normal" polar i.e. it accretes via Roche lobe overflow. Furthermore, the ZTF phase folded light curves (Appendix A) confirm this behaviour. Unfortunately, as mentioned above, the sole spectrum is of very low S/N, as it was taken just prior to the start of the first light curve, when the source was in low accretion state, and cannot be used to estimate the WD magnetic field strength via detection of cyclotron humps. Finally, we note that the ZTF data reveal rather varied time dependent light curve pulse shapes, indicative of perhaps unstable accretion geometry or changing accretion rate over hundreds of days.

\subsection{ZTF J2220+0721}

Another eclipsing source (P=0.140 d) from \citet{Brown2023} that shows a strong, isolated cyclotron hump at around 4800\AA\ superimposed on what resembles a late type M dwarf spectrum with characteristic molecular band heads. The strong modulation seen in the g band of ZTF light curves (Appendix A) can be explained by the orbital modulation of this cyclotron hump. The circular polarisation behaviour is peculiar, as the cyclotron hump shows a double peaked structure (Fig \ref{fig2220}). The source appears similar to ZTF0146+4914 \citep{Guidry2021,Hakala2022}, as it exhibits a similar strong isolated cyclotron hump in the blue. No other clear humps, although there is a slight concentration of negative circular polarisation at around 9500\AA. If these two were the neighbouring harmonics, then the field could be as high as $\sim$110 MG. There is no indication of H$_\alpha$, or any other emission lines. Even if there is some indication that the brightness of the molecular band features (at around 7500\AA\ ) in the trailed spectrum follows the orbital modulation of the cyclotron hump at 4800\AA\ , there is no circular polarisation, and the overall spectrum suggests they are only molecular bandheads. The ZTF light curves (Appendix A) show remarkable stability suggesting a very stable long term accretion rate. This, together with the above mentioned other properties, make this source likely our strongest case for a LARP.

\subsection{ZTF J2353+4153}

The 0.174 d period source was classified as an eclipsing magnetic system by \citet{Brown2023} and it appears to show two clear cyclotron harmonics in the circularly polarised spectrum, which can be produced from a 54 MG field (Fig \ref{fig2353spol}). The red vertical markers show the expected centres of 3rd and 4th harmonics from such field (using eqn. 1)
. The blue markers give the 
$\pm$1MG model hump locations. These are not easily reproduced with any other field strengths (i.e. assuming different pair of harmonics). The circular polarisation features short of 4600\AA\ are likely not real, as the spectral extraction becomes very noisy in that region.

We have carried out cyclotron emission mapping on the WD surface, as in Paper I, by fitting the photometry and circular polarisation curves using cyclotron models. In particular, we generated a bespoke cyclotron model for the NOT skycontrast filter (\#92) transmission curve  using the constant temperature and constant $\Lambda$ cyclotron emission models and code by \citet{1992A&A...256..498W}, \citet{1993A&A...280..169W} and \citet{1996A&A...310..526R}. For the models we employed B=54MG, T=10keV and the plasma parameter log $\Lambda$=5.0. The results of our modelling are also shown in Fig \ref{ztf2353model}. Our modelling results suggest accretion both on positive and negative poles, which are located approximately on the opposite sides of the WD. However, it appears a pair of positive and negative poles are required. This further suggests that the WD magnetic field geometry is likely not a simple dipole. Very similar results were obtained in Paper I for a single pole accretor SDSS J2229+1853 (Fig 9. in Paper I). Furthermore, the eclipse profile modelling of SDSS J030308.35+005444.1 \citep{Parsons2013} also came up with similar accretion geometry (or accreting pole locations, where two poles, facing the secondary star, are located in the different WD hemispheres). Obviously \citet{Parsons2013} were not able to determine the magnetic field signs at those location, as they only possessed photometric data.

The optical light curves of SDSS J2229+1853 (Paper I) and ZTF J2353+4153 both show distinct lack of any flickering, suggestive of homogeneous accreting matter that could be provided by the donor wind, rather than via Roche lobe overflow.

\begin{table*}[ht!]
  \begin{tabular}{llll}
    \hline
Source & WD B-field & Notes & Type ID\\
    \hline\hline
    
ZTF J0056+4926 & $\sim$40 MG & Strong variable $H_\alpha$, cyclotron humps in circ.pol. spectrum. & LARP\\    

ZTF J0220+6303 &  & Curious 2-comp. spectrum with no clear circ. pol. & ?\\

ZTF J0406+0958 &  & Ellipsoidal modulation with offset eclipse phase. No circ. pol. & ? \\ 	

ZTF J0504+2145 &  & A blue spectrum with circ. pol. in the red tail. & Novalike CV (IP?) \\ 

ZTF J0753+3203 &  & Detached eclipsing binary & MS+WD\\ 

ZTF J1520+3453 &  & Late type spectrum, some $H_\alpha$ emission. No clear circ. pol. & ? \\

ZTF J1737+4013 &  & Modulated circ. pol. + accretion mode changes & Polar\\

ZTF J2220+0721 & $\sim$110 MG? & Very strongly modulated cyclotron hump at 4850\AA. No clear 2nd hump. & LARP \\

ZTF J2353+4153 & 54 MG & Clear cyclotron humps. & LARP\\

\hline
  \end{tabular}
  \caption{The summary of observed properties, together with the measured magnetic fields, of the candidate systems.}
  \label{summaryproperties}
\end{table*}

\section{Discussion}
Low accretion-rate polars (LARPs), often associated with the class of pre-polars or wind-accreting magnetic post-common-envelope binaries, occupy a key evolutionary niche between detached magnetic white dwarf binaries and fully developed, Roche-lobe–overflowing cataclysmic variables (CVs). 

LARPs represent the extreme low end of the mass-transfer distribution in such systems, with accretion likely powered by the stellar wind of the secondary rather than sustained Roche-lobe overflow (e.g. \citealt{Schwope2002}; \citealt{Schmidt2005}).

Moreover, because their optical emission is typically dominated by discrete cyclotron harmonics with minimal contamination from a bright accretion stream or disc, wind-accreting magnetic CVs provide one of the cleanest observational windows onto magnetically channelled accretion physics. Cyclotron spectroscopy and polarimetry directly constrain magnetic field strengths, plasma temperatures, and accretion geometry (e.g. \citealt{Wickramasinghe2000}). In this respect, LARPs serve as laboratories for studying accretion shocks and radiative transfer in strong magnetic fields under low-density conditions, complementing the higher-accretion-rate regime observed in classical polars.

In this paper, we have presented optical photopolarimetry and spectropolarimetry of a sample of nine eclipsing LARP candidates selected from the \citealt{Bellm2019} Zwicky Transient Facility (ZTF) survey on the basis of their peculiar light curve morphologies, colours, and/or eclipse profiles. Time-resolved polarimetry remains the most robust diagnostic for identifying magnetic accretion, as significant circular polarisation unambiguously signals cyclotron emission from magnetically confined accretion regions.

We find that three of these systems are very likely bona fide LARPs and one is a “normal” polar undergoing Roche-lobe overflow. For the remaining systems, apart from one novalike CV, the results are either inconclusive or inconsistent with accretion onto a magnetic WD at the levels expected for LARPs. This can be due to the non-visibility of the accretion region during the observations (in case of partial orbital phase coverage in some systems). It is worth noting that in most cases also the variability due to the donor star can be ruled out based on the ZTF light curves, leaving the explanation unresolved.

A summary of the properties and classifications of the individual systems is provided in Table~\ref{summaryproperties}. The confirmation rate of $\sim$1/3 underscores both the promise and the limitations of variability-based pre-selection from wide-field optical surveys.
Looking forward, we expect that many more LARP candidates will be identified from ongoing and forthcoming ground-based surveys such as ZTF and, in the near future, the Rubin Observatory LSST. The depth, cadence, and multi-band coverage of these facilities will dramatically increase the discovery space for faint, low-accretion magnetic binaries. However, our results demonstrate that photometric selection alone is insufficient to efficiently isolate true LARPs. Refinement of selection criteria, potentially incorporating colour evolution across orbital phase, machine-learning classification of light curve morphology, utilising the apparent lack of flickering, cross-matching with X-ray catalogues, and spectroscopic indicators of cyclotron humps. will be essential to improve yield.
Systematic, time-resolved polarimetric follow-up remains critical for unambiguous classification. In parallel, incorporating confirmed LARPs into population studies will allow more rigorous comparison with theoretical predictions, constraining both the space density of magnetic post-common-envelope binaries and the efficiency of AML mechanisms at low mass-transfer rates. Expanding the confirmed sample will also clarify whether LARPs constitute a transient evolutionary phase, a long-lived stable configuration, or a heterogeneous mixture of systems at different evolutionary stages.

\section{Conclusions}

In conclusion, our study has revealed three new LARP systems. LARPs represent a pivotal population for understanding the magnetic and secular evolution of close binaries. By combining wide-field time-domain surveys with targeted polarimetric follow-up, we are now in a position to move from serendipitous discovery toward a statistically meaningful census. Such a census will not only refine models of CV evolution but also deepen our understanding of magnetically controlled accretion under some of the lowest mass-transfer conditions accessible in interacting binaries.

\begin{acknowledgements}
      SGP acknowledges support by the Science and Technology Facilities Council (grant ST/B001174/1). This project has received funding from the European Research Council (ERC) under the European Union’s Horizon 2020 research and innovation programme (Grant agreement No. 101020057).
      We cannot thank enough the NOT staff that carried out the service observations braving
      both the raging pandemic and the Volcanic eruption on La Palma at the time of observations. Based on observations made with the Nordic Optical Telescope, owned in collaboration by the University of Turku and Aarhus University, and operated jointly by Aarhus University, the University of Turku and the University of Oslo, representing Denmark, Finland and Norway, the University of Iceland and Stockholm University at the Observatorio del Roque de los Muchachos, La Palma, Spain, of the Instituto de Astrofisica de Canarias. The NOT data were obtained under program ID P64-003. The data presented here were obtained with ALFOSC, which is provided by the Instituto de Astrofisica de Andalucia (IAA) under a joint agreement with the University of Copenhagen and NOT. 
      This work has made use of data from the European Space Agency (ESA) mission
{\it Gaia} (\url{https://www.cosmos.esa.int/gaia}), processed by the {\it Gaia}
Data Processing and Analysis Consortium (DPAC,
\url{https://www.cosmos.esa.int/web/gaia/dpac/consortium}). Funding for the DPAC
has been provided by national institutions, in particular the institutions
participating in the {\it Gaia} Multilateral Agreement.\\ 

Finally, we would like to extend our thanks to Ilaria Caiazzo, whose comments as the referee, certainly improved the article in several ways.
\end{acknowledgements}

%


%

\clearpage
\let\cleardoublepage\clearpage

\appendix

\section{ZTF light curves}

\begin{figure*}
  \includegraphics[width=0.97\textwidth]{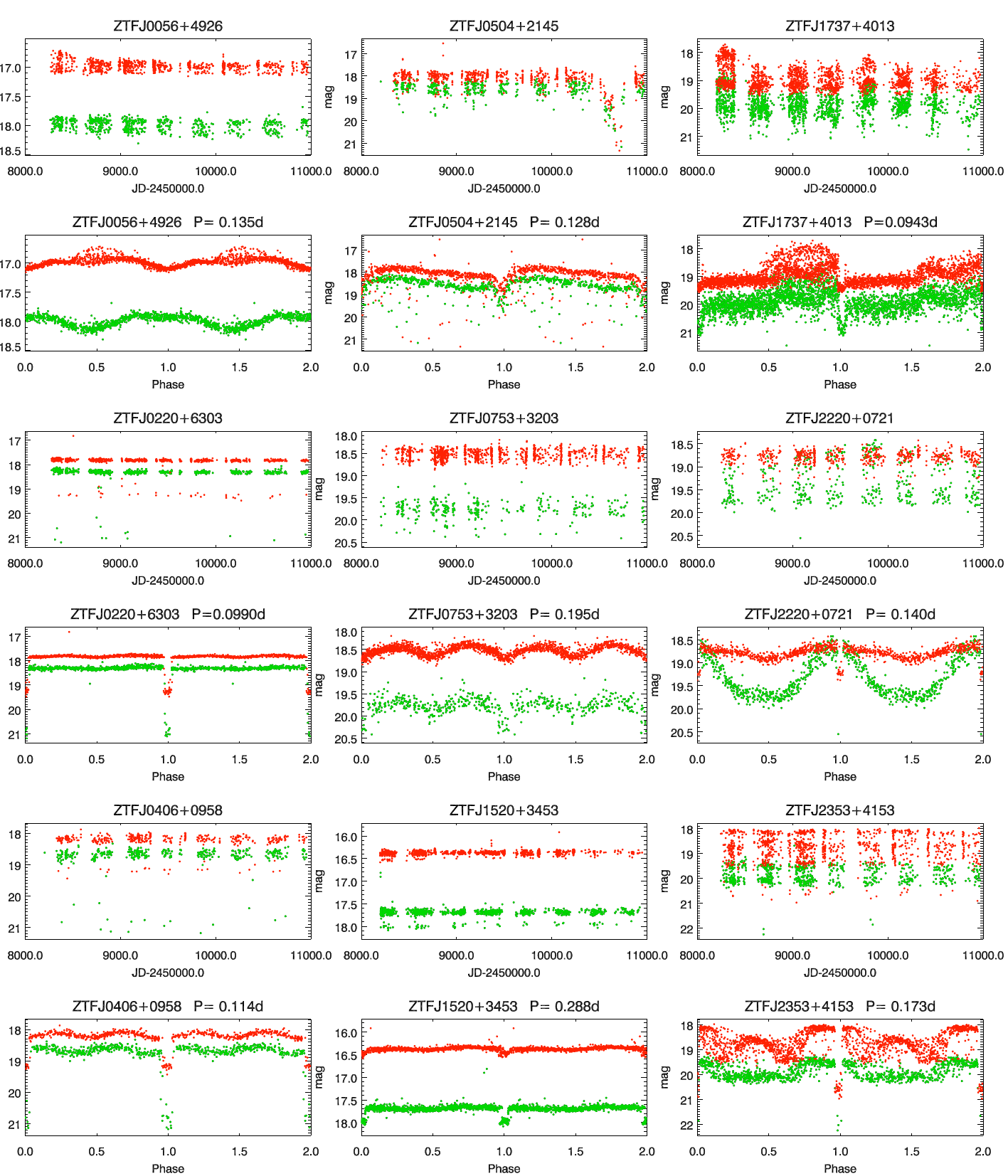}
    \caption{The ZTF light curves of all candidate sources. For each source, we have plotted the unfolded light curves on top and phase folded light curves below. The data are plotted twice (from 0.0 to 2.0) for clarity. The g band data is marked with green and r band data with red dots. The red curves are above the green curves for every plot.}
    \label{ztffig}
\end{figure*}
%



\end{document}